\documentclass[prl,twocolumn,showpacs,superscriptaddress,amsmath,amssymb]{revtex4}
\usepackage{graphicx}
\usepackage{dcolumn}
\usepackage{bm}

\def\kbar{\protect\@kbar}
\def\@kbar{%
\relax \bgroup
\def\@tempa{\hbox{\raise.73\ht0
\hbox to0pt{\kern.25\wd0\vrule width.5\wd0
height.1pt depth.1pt\hss}\box0}}%
\mathchoice{\setbox0\hbox{$\displaystyle k$}\@tempa}%
{\setbox0\hbox{$\textstyle k$}\@tempa}%
{\setbox0\hbox{$\scriptstyle k$}\@tempa}%
{\setbox0\hbox{$\scriptscriptstyle k$}\@tempa}%
\egroup}

\begin{document}

\title{Observation of persistent flow of a Bose-Einstein condensate in a toroidal trap}

\author{C. Ryu}
\affiliation{Atomic Physics Division, National Institute of Standards and Technology, Gaithersburg, Maryland 20899-8424, USA}
\affiliation{Joint Quantum Institute, NIST and University of Maryland, College Park, Maryland 20742, USA}
\author{M. F. Andersen}
\altaffiliation[Permanent Address: ]{Department of Physics, University of Otago, Dunedin 9016, New Zealand}
\affiliation{Atomic Physics Division, National Institute of Standards and Technology, Gaithersburg, Maryland 20899-8424, USA}
\author{P. Clad\'e}
\affiliation{Atomic Physics Division, National Institute of Standards and Technology, Gaithersburg, Maryland 20899-8424, USA}
\author{Vasant Natarajan}
\altaffiliation[Permanent Address: ]{Department of Physics, Indian Institute of Science, Bangalore 560012, India}
\affiliation{Atomic Physics Division, National Institute of Standards and Technology, Gaithersburg, Maryland 20899-8424, USA} 
\author{K. Helmerson}
\affiliation{Atomic Physics Division, National Institute of Standards and Technology, Gaithersburg, Maryland 20899-8424, USA}
\affiliation{Joint Quantum Institute, NIST and University of Maryland, College Park, Maryland 20742, USA}
\author{W. D. Phillips}
\affiliation{Atomic Physics Division, National Institute of Standards and Technology, Gaithersburg, Maryland 20899-8424, USA}
\affiliation{Joint Quantum Institute, NIST and University of Maryland, College Park, Maryland 20742, USA}

\date{\today}

\begin{abstract}

We have observed the persistent flow of Bose-condensed atoms in a toroidal trap. The flow persists without decay for up to 10 s, limited only by experimental factors such as drift and trap lifetime. The quantized rotation was initiated by transferring one unit, $\hbar$, of the orbital angular momentum from Laguerre-Gaussian photons to each atom. Stable flow was only possible when the trap was multiply-connected, and was observed with a BEC fraction as small as 20\%. We also created flow with two units of angular momentum, and observed its splitting into two singly-charged vortices when the trap geometry was changed from multiply- to simply-connected.

\end{abstract}


\maketitle
One of the most remarkable properties of macroscopic quantum systems is the phenomenon of persistent flow. In a superconductor, persistent flow is electrical current without resistance:  Current in a loop of superconducting wire will flow essentially forever. In a superfluid \cite{leggett1999} such as liquid helium below the lambda point, the frictionless flow allows persistent circulation in a hollow toroid.

A Bose-Einstein condensate (BEC) of an atomic gas also exhibits superfluidity. 
Experiments have confirmed the superfluid behavior by demonstrating a critical velocity below which a laser beam could be moved through the gas without causing excitations \cite{raman1999}, and irrotational flow through the creation of vortices \cite{dalibard2000} and vortex lattices \cite{ketterle2001} in both rotating and non-rotating traps.
While such vortex motion can be long-lived even in non-rotating traps \cite{haljan2001}, they are always unstable. A toriodal trap would allow stability, because it costs too much energy for the vortex core to move from the center of the torus, where the density is zero, through the high density atomic cloud \cite{leggett2001, bloch1973, abraham2001}. Hence, until now, such stable, persistent flow has not been observed. 

Persistent flow of a BEC, in addition to being a striking demonstration of superfluid behavior, would facilitate the understanding of the nature of the critical velocity in superfluid liquid helium \cite{reppy1967}, which is still not well understood. 
Persistent flow of a BEC could also be used to understand the fundamental relationship  between superfluidity and Bose condensation, especially in one- and two-dimensional systems. For example, a strongly-interacting 1-D gas is predicted to show superfluidity \cite{kagan2000}, although such a system does not Bose condense.

In order to create stable, persistent flow, we confine our BEC
in a toroidal (multiply-connected) trap, such that the density of the condensate is continuous throughout the torus and its phase is uniform.
There have been numerous proposals of such traps for BECs \cite{wright2001, henne2006, mabuchi2004, schmi2006}, with experiments demonstrating
the loading of a condensate into a large diameter ring-shaped trap \cite{gupta2005, riis2006} 
in such a way that when the atoms subsequently spread out to fill the ring and they are no longer in the many-body ground state. A true BEC in a multiply connected geometry was realized earlier in an experiment to measure the onset of dissipation by scanning a blue-detuned laser beam through a condensate confined in a cigar-shaped, harmonic, magnetic trap \cite{raman1999, onofrio2000}.

Our toroidal trap also uses a blue-detuned laser beam to make a repulsive potential barrier in the middle of a harmonic, magnetic trap. 
The potential from our harmonic trap and the Gaussian laser beam, propagating in the $x$ direction with waist $w_{0}$ \cite{footnote1}, is $\frac{1}{2} m ({\omega_{x}}^2 x^2 + {\omega_{y}}^2 y^2 + {\omega_{z}}^2 z^2) + V_{0} \exp(-2(y^2+z^2)/w_{0}^{2})$, where $m$ is the atomic mass and $\omega_{x,y,z}$ are the harmonic trapping frequencies along $x, y$ and $z$, respectively. $V_{0}$ is the maximum optical potential, proportional to the laser intensity. 
The plug beam repels atoms from the trap center creating a BEC density distribution with a hole in the center, if the chemical potential, $\mu$, is less than $V_{0}$. Hence, this trap geometry can be conveniently changed from simply-to multiply-connected by increasing the laser power. Similarly, changing the intensity and size of the plug beam changes the radii and shape of the toroidal trap.

Radio frequency (rf) induced evaporation is used to cool a cloud of Na atoms in the $|3S_{1/2}, F=1, m_F =-1\rangle$ state confined in our triaxial, time orbiting potential (TOP) magnetic trap \cite{kozuma1999} to just above the BEC transition temperature. The TOP trap is then expanded (during 0.5 s) to the final trapping frequencies ($\omega_{x, y, z}/2\pi = 25, 36$ and $51$ Hz, respectively, with gravity along $z$). Simultaneously, the plug beam (wavelength $532$ nm, $w_{0} = 8 \mu$m) power is increased linearly from zero to $\approx 600 \mu$W. (A single-mode optical fiber is used to provide good mode quality and directional stability.) A final rf evaporation (over 5.6 seconds) in the combined (toroidal) trap while the plug beam power was reduced to the final value, depending on the experiment, resulted in a nearly pure BEC of about  $2-5  \times 10^{5}$ atoms. Typically, the rf is left on, tuned slightly higher than the final rf evaporation frequency, to suppress any subsequent heating.

Fig.\ 1(a) shows the combined trapping potential in the $y$-$z$ plane at $x = 0$. Since our TOP trap is not rotationally symmetric, neither is the combined potential. Hence the toroidal potential has (two) local minima (red region in Fig.\ 1(a)). However, a continuous, toroidal-shaped BEC (see Fig.\ 1(b)), characterized by a single order parameter, can still be made in this trap since the variations in the depth around the potential is about $\mu /3$ \cite{footnote2}.

\begin{figure}
\includegraphics[angle=0,scale=0.17]{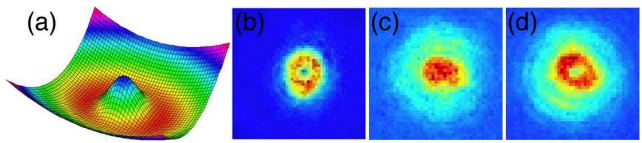}
\caption{(a) Toroidal trap from the combined potentials of the TOP trap and Gaussian plug beam. (b) {\it In-situ} image of a BEC in the toroidal trap. (c) TOF image of a non-circulating BEC released from the toroidal trap. (d) TOF image of a circulating BEC, released after transfer of $\hbar$ of OAM.}
\label{schematic}
\end{figure}

We create circulation in the toroidal trap by transferring the orbital angular momentum (OAM) of a Laguerre-Gaussian (LG) laser beam to the atoms using the experimental procedure described in \cite{oamprl}. The transfer of OAM is accomplished by a resonant, two-photon, stimulated Raman process with a Gaussian (G) laser beam propagating along $x$ and a LG beam, carrying $\hbar$ of OAM, propagating along $-x$. Atoms that absorb a photon from one beam and stimulatedly emit a photon into the other beam acquiring both the linear momentum (LM) and OAM difference of the beams \cite{oamprl}, which in this case is $2\hbar k$ and $\hbar$, respectively ($k$ is the magnitude of the photon wavevector). To achieve OAM transfer without net LM transfer, we initially transfer all of the atoms to the $2\hbar k$ LM state using counterpropagating G beams \cite{kozuma1999} along $x$, and then rapidly (within $\approx 50 ~\mu$s) transfer the atoms back to the zero LM state using the G and LG beam pair. The entire process results in about 50\% of the atoms with $\hbar$ of OAM. To remove the atoms moving at $2 \hbar k/m$ (6 cm/s) we wait for a quarter of the trap oscillation period ($\pi/2\omega_{x}$)
and then optically pump these atoms into the untrapped $F=2$ hyperfine states using a spatially localized laser beam \cite{oamprl}. The whole process takes only 7 ms. Excitations from the OAM transfer thermalizes in a few hundred ms and are partially removed by the rf, which was kept on since forming the BEC.

Fig.\ 1(c) shows the time-of-flight (TOF) expansion image of the non-rotating BEC taken 18 ms after instantaneously ($<100\ \mu$s) extinguishing the toroidal trap. The initial hole in the BEC (see Fig.\ 1(b)) is filled in during expansion \cite{stringariRMP}. In Fig.\ 1(d), we show the corresponding TOF image of a circulating BEC after transfer of $\hbar$ of OAM. The hole is clearly visible providing a simple way to detect circulation \cite{oamprl, stringariPRA}.

In order to investigate persistent flow in a condensate, we measured the decay of the circulation as a function of time in the toroidal trap. For this experiment, the plug beam power was 113 $\mu$W ($V_{0}/h = 3.6$ kHz) and the average number of atoms was about $2.5\times 10^{5}$. In Fig.\ 2, we plot the probability of observing the circulating state \cite{footnote3} as a function of time held in the trap. The circulation in the toroidal trap persists without decay for up to 10 s. In contrast, the circulation in only the TOP trap (without the plug beam) decays in about 0.5 s \cite{footnote3.5}. 

The persistence of the flow, which can be understood from the energy argument given above, can also be understood as the flow velocity being less then the superfluid critical velocity, which for a uniform system is given by the sound velocity $v_{s}=\sqrt{\mu/m}$ \cite{stringariPRA}. For our typical experimental conditions ($\mu/h \approx 0.5$ kHz), where the inner diameter $r_{0}$ ($\approx 10\ \mu$m) of the torus is very much greater than the healing length $\hbar/(\sqrt{2}mv_{s})$ \cite{stringariRMP}, the flow velocity for unit winding number, $v_f=\hbar/m r_{0}$, is $0.29$ mm/s, while $v_{s} = 2.9$ mm/s where the density is highest. The flow velocity is much less than the sound velocity and therefore less than the critical velocity, so the stability condition is satisfied.

\begin{figure}
\includegraphics[angle=0,scale=0.60]{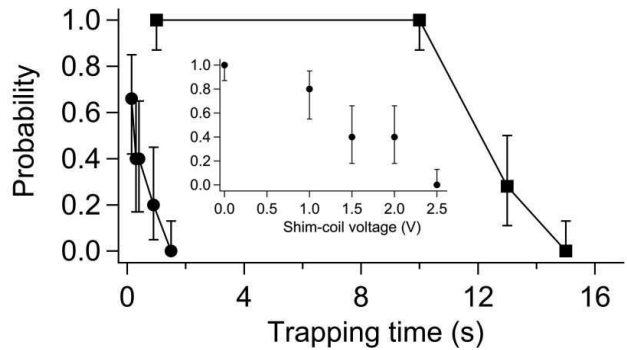}
\caption{Survival probability of circulation as a function of trapping time in the TOP trap without the plug beam (circles) and with the plug beam (squares). Each data point is an average of 5-10 measurements. (inset) Survival probability for 0.5 s trapping time, as a function of the voltage applied to external shim-coils to displace the TOP trap center. 
} 
\label{picture}
\end{figure}

In Fig.\ 2(inset) we show the survival probability of the flow as the relative position of the plug beam and the TOP trap center is varied. In this case, we first establish circulation in the toroidal trap, waiting 500 ms for the flow to stabilize, and then increase the voltage to external magnetic coils (during 200 ms) to displace the TOP trap center ($0.8\ \mu$m/V). The atoms held in the misaligned, plugged trap for 500 ms were then released for TOF imaging. As the flow becomes spatially constricted due to misalignment, the local flow velocity at the constriction has to increase for constant circulation of atoms, and the critical velocity decreases due to the decrease in local density. If the local flow velocity exceeds the critical velocity, the flow will no longer be superfluid resulting in the observed decay of the flow. 

A similar misalignment is responsible for the decay of persistent flow after 10 s, as seen in Fig.\ 2. This misalignment is due to a relative drift ($0.4\ \mu$m/s) between the position of the plug beam and the center of the TOP magnetic trap, which drifts due to thermal cycling during each realization of the experiment. We substantially compensate for this drift  by applying a linear voltage ramp to the external coils to generate a linear, temporal shift in the location of the TOP trap center, but the residual drift eventually (after 10 s) leads to a loss of flow. Thus, it appears that if the drift could be completely eliminated, the circulation in the toroidal trap would indeed be  {\it persistent}; however, the vacuum limited lifetime of our trapped atoms is about 15 s.

An alternative view of the persistent circulation of atoms in the toroidal trap is that of a macroscopic rotational or vortex state pinned by the potential created by the plug beam \cite{footnote4}. We investigated the ability of the plug beam to pin the vortex state by measuring the survival probability of the circulation after a trapping time of 2 s, for different values of the plug beam power. Fig.\ 3 shows the survival probability versus the maximum height of the optical potential, $V_0$ (calculated from the measured plug beam power and waist) scaled by the chemical potential (calculated from the measured number of atoms and the toroidal trapping potential). 
The survival probability increases above $V_{0}/\mu = 1$, which corresponds to the condition of the trap being multiply- vs. simply-connected. Eventually, when the height of the pinning potential is sufficiently large, the vortex is stable. The actual value of $V_{0}/\mu$ has an uncertainty of about 10\% due to fluctuations in the plug laser power, and there is a systematic error of about 30\% due to the uncertainty in the measurement of the plug beam waist.

\begin{figure}
\includegraphics[angle=0,scale=0.45]{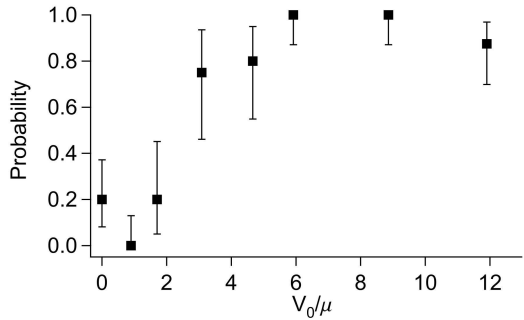}
\caption{Survival probability for a 2 s trapping time, as a function of the maximum height of the plug beam optical potential, $V_{0}$, (normalized to the chemical potential, $\mu$).}
\label{quantum}
\end{figure}

One of the interesting aspects of persistent flow is its stability at finite temperatures. To investigate this, we varied the thermal fraction of the cloud by varying the end point frequency of rf evaporation. (The rf was then turned off after evaporation.) We then measured the survival probability of the circulation after 2 s of trapping time for various thermal fractions. The BEC fraction is determined from a measurement of the number of atoms and the thermal fraction \cite{footnote5}; however, because of additional heating in the combined optical and magnetic trap (possibly due to relative motion), we observe that the BEC fraction decreases during the trapping time. Hence, the estimated BEC fraction from the TOF images is the final fraction, and the initial fraction is probably higher \cite{footnote6}. As seen in Fig.\ 4, the flow survives even for a BEC fraction as small as 20\%. The persistence of circulation for such a small BEC fraction is interesting, since the existence of even a small thermal fraction causes dissipation of a vortex in a simply-connected trap \cite{raman2002}.

\begin{figure}
\includegraphics[angle=0,scale=0.55]{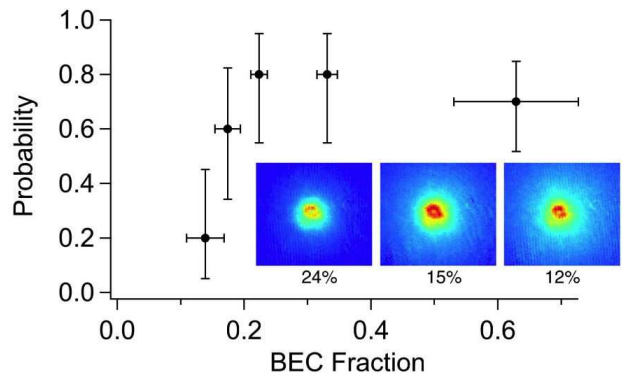}
\caption{Survival probability of a circulating BEC for a 2 s trapping time, as a function of the BEC
fraction. (inset) Three TOF images of a circulating BEC for BEC fractions of 24, 15 and 12\%.} 
\label{temperature}
\end{figure}

We can also create higher circulation in the toroidal trap by transferring higher angular momentum to the atoms. Such flow can be stable, even under conditions for which it is unstable in a simply-connected trap. To demonstrate this, we used a larger waist ($w_0=15\ \mu$m) plug beam, to reduce the flow velocity of the circulating cloud, so that the stability criterion can be satisfied. In order to keep the trap sufficiently flat azimuthally with the larger plug \cite{footnote2}, we made the TOP trap closer (10\% variation in the radius) to cylindrically symmetric \cite{cornellsphericalTOP}. To generate higher circulation, we used an initial LG/G pulse to transfer the atoms into a state with OAM of $\hbar$ and LM of $2\hbar k$. We then used another LG/G pulse, but with a LG beam having opposite OAM, to transfer atoms back to the zero LM state with an additional OAM of $\hbar$. The result is a trapped cloud, of $\approx$50\% of the initial atoms, circulating with $2 \hbar$ of angular momentum.

Fig.\ 5 shows images, after 20 ms TOF, of the circulating cloud with two units of angular momentum for different hold times in just the TOP trap. The flow is initially stabilized in the toroidal trap for 0.5 s, and then the plug beam power is ramped down in 30 ms (to avoid creating excitations). In Fig.\ 5(a) (4 ms hold time in the TOP trap) there is a single, larger hole in the center due to the higher angular momentum \cite{oamprl, stringariPRA}. To confirm that this circulating state corresponds to a doubly-charged vortex, we allowed the cloud to evolve in the TOP trap for longer times. Figs.\ 5(b) and (c) show the splitting of the doubly-charged vortex into two singly-charged vortices.

\begin{figure}
\includegraphics[angle=0,scale=0.55]{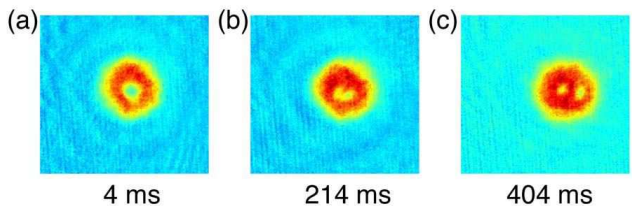}
\caption{ TOF pictures of doubly-charged flow, initially stabilized for 0.5 s in the toroidal trap, for different holding times (4, 214 and 404 ms) in just the TOP trap.}
\label{doublevortex}
\end{figure}

Superfluid flow in a ring geometry raises interesting new possibilities. For example, the splitting times of multiply-charged vortices \cite{ketterle2004, splittingtheory}, especially for odd winding number, could be investigated. With the addition of a quantum tunnel barrier, acting as a Josephson junction, the analog of a superconducting quantum interference device (SQUID) could be realized for a gas of atoms \cite{andersonAPHID}. 
Furthermore, there is the open question regarding the relation between superfluidity and Bose condensation, especially in lower dimensions. 
Modifications should allow creation of a strongly-interacting 1-D gas in a tightly-confined ring to study its superfluid properties.


\begin{references}
\bibitem{footnote1} For our experiments, the Rayleigh length of the laser beam is more than a factor of ten longer than the extent of the BEC, therefore we approximate the laser beam as having a uniform radial extent, given by the waist, $w_{0}$, over the extent of the BEC in the $x$ direction.
\bibitem{footnote2} This variation in the minimum of the toroidal potential increases when the radius of the toroidal trap is increased by increasing the waist of the plug laser. In order to keep these variations less than the chemical potential, the TOP trap can be made more symmetric \cite{cornellsphericalTOP}. Alternatively, the plug beam can be made elliptical with an aspect ratio matching the TOP trap.
\bibitem{footnote3} The probability of observing the circulating state was obtained by looking for the presence of the central hole in the image of the distribution of atoms after a sufficient TOF period, and then taking the ratio of the number of such images divided by the total number of images for multiple realizations of the experiment.
\bibitem{footnote3.5} The observed vortex decay corresponds to the absence of a hole in the TOF image of the atoms. We can not distinguish whether this is due to the vortex leaving the trap, or rotating in the trap such that the vortex core is no longer along the axis of our imaging beam.
\bibitem{footnote5} To determine 
the thermal fraction, we fit to each TOF image (Fig.\ 4(inset)) a Gaussian to only the wings of the distribution, which presumably corresponds to the broad, thermal (non-condensed) component.
\bibitem{footnote6} For example, we observe that an initial 10\% BEC fraction decreased to 4\% after 2 seconds.
\end{references}
\end{document}